\documentclass{aastex63}

\usepackage{amsmath}

\hypersetup{linkcolor=red,citecolor=blue,filecolor=green,urlcolor=magenta}

\submitjournal{RNAAS}

\begin{document}

\shorttitle{JWST Parallax}
\shortauthors{Ngeow et al.}

\title{Demonstrating the Concept of Parallax with James Webb Space Telescope}

\correspondingauthor{C.-C. Ngeow}
\email{cngeow@astro.ncu.edu.tw}

\author[0000-0001-8771-7554]{Chow-Choong Ngeow}
\affil{Graduate Institute of Astronomy, National Central University, 300 Jhongda Road, 32001 Jhongli, Taiwan}

\author[0000-0003-0871-4641]{Harsh Kumar}
\affil{Indian Institute of Technology Bombay, Powai, Mumbai 400076, India}

\author[0000-0002-6112-7609]{Varun Bhalerao}
\affil{Indian Institute of Technology Bombay, Powai, Mumbai 400076, India}

\begin{abstract}
We measured the parallax of the James Webb Space Telescope based on near simultaneous observations using the Lulin One-meter Telescope and the GROWTH India Telescope, separated at a distance of $\sim 4214$~km. This serves a great demonstration for the concept of parallax commonly taught in introductory astronomy courses.

\end{abstract}


\section{Introduction}

The James Webb Space Telescope (JWST) was launched on 25 December 2021, and after a journey of about a month, JWST arrived at its final destination -- the second Lagrange point (L2) of the Sun-Earth orbit. The L2 is located at $\sim 1.5 \times 10^6$~km away from the Earth, implying it is possible to measure the parallax of JWST from two distant sites on Earth. For example, two sites separated by 100~km will be able to measure a parallax of $\sim 6.88\arcsec$. Therefore, JWST on L2 provides a great opportunity to demonstrate parallax, an important astronomical concept to be taught in introductory astronomical cources. In this work, we perform near simultaneous observations on JWST from two sites, and demonstrated that it is possible to measure parallax, and hence the distance from Earth, for JWST. This will be of great interest for educational purposes. 

\section{The Near Simultaneous Observations and the Distance of JWST}

\begin{figure*}
  \epsscale{1.2}
  \plotone{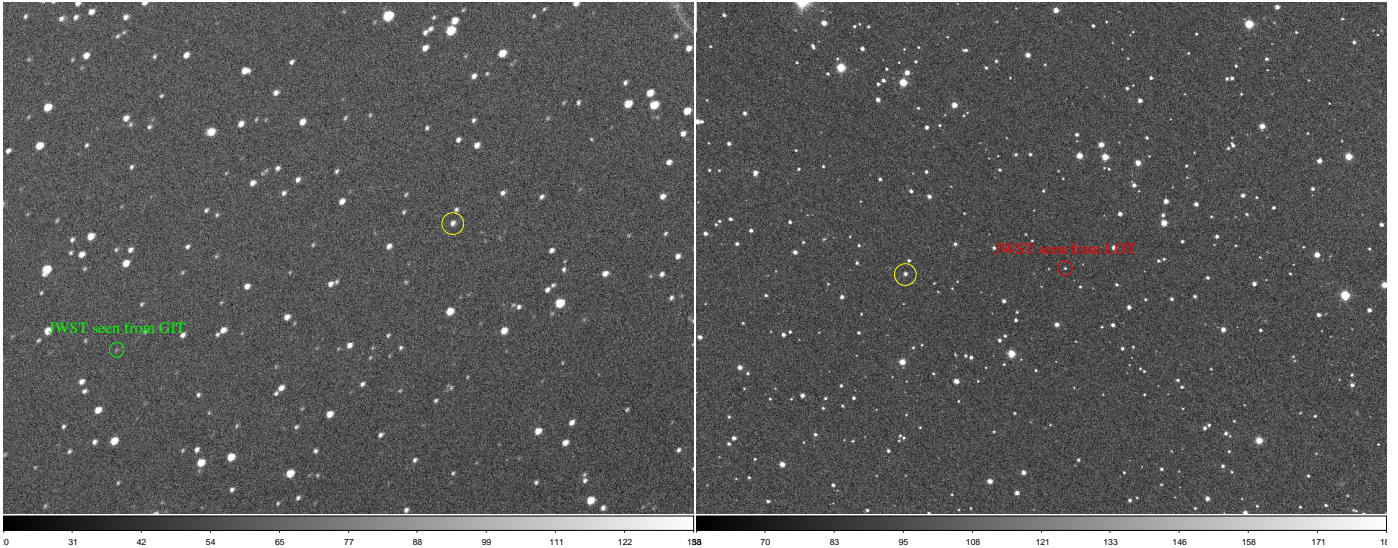}
  \caption{Near simultaneous images taken from GIT (left panel) and LOT (right panel). Locations of JWST on these images, based on the information taken from \url{https://theskylive.com/jwst-info}, are marked. The yellow circles on both images represent a randomly chosen reference star to guide the eyes for the relative position of JWST on these images. Both images were reduced in a standard manner (bias and dark subtracted, and flat-fielded). Astrometric refinement was done using the {\tt astrometry.net} package \citep{lang2010}. The measured JWST coordinates are 07h26m40.0s, 09d57m04.6s for GIT and 07h26m03.0s, 09d59m30.0s for LOT.} \label{fig}
\end{figure*}

A coordinated observation was carried out on 08 February 2022 using the Lulin One-meter Telescope (LOT, located in Taiwan; $120^\circ 52\arcmin 25\arcsec E,\ 23^\circ 28\arcmin 07\arcsec N$) and the 0.7-m GROWTH India Telescope (GIT, located in India; $78^\circ 57\arcmin 55.1\arcsec E,\ 32^\circ 46\arcmin 44.1\arcsec N$). A sequence of 17 images were taken from both telescopes starting at UT 14:52, and we identified a pair of images that were closest in time (UT 15:05:38 and 15:05:22 for GIT and LOT, respectively). JWST was clearly detected on both images, as shown in Figure \ref{fig}. We measured the angular separation of JWST on both images, which was found to be $\sim 566.0\arcsec$, or a parallax of $\sim 283.0\arcsec$. Given that the distance between GIT and LOT is $\sim 4214.17$~km, our measured parallax of JWST translates to an approximate distance of $\sim 1.5358 \times 10^6$~km.

We refine this calculation by correcting for two factors. First, the straight-line distance between GIT and LOT is shorter than the distance along the surface. We use the \texttt{EarthLocation} feature in astropy to define the two observatory locations, and find that the direct distance between them is 4142~km. Second, the line joining the observatories was not exactly perpendicular to the line of sight to JWST, but has an angle of 79\degr.6. 
With these values, the corrected distance is $1.4849\times 10^6$~km. Our astrometric uncertainty of about 0\arcsec.08 yields a distance uncertainty of about 200~km.

\section{Conclusion}

In this work, we demonstrated that parallax can be measured for JWST based on two distant sites on Earth, which can be of a great interest for teaching the concept of parallax. Some education-friendly animations are available at \url{https://sites.google.com/view/growthindia/outreach/spotting-jwst}. The pair of images are also available at the same URL, which can be used in various educational purposes (for examples, measuring positions and parallax of JWST, distance calculation, etc).

\acknowledgments

We thank the observing staff at Lulin Observatory, C.-S. Lin, H.-Y. Hsiao, and W.-J. Hou, to carry out the requested observations. This publication has made use of data collected at Lulin Observatory, partly supported by MoST grant 109-2112-M-008-001. The GROWTH India Telescope (GIT) is a 70-cm telescope with a 0.7-degree field of view, set up by the Indian Institute of Astrophysics (IIA) and the Indian Institute of Technology Bombay (IITB) with funding from DST-SERB and IUSSTF. It is located at the Indian Astronomical Observatory (Hanle), operated by IIA. We acknowledge funding by the IITB alumni batch of 1994, which partially supports operations of the telescope. 
This research made use of Astropy,\footnote{\url{http://www.astropy.org}} a community-developed core Python package for Astronomy \citep{astropy2013, astropy2018}.

\facility{LO:1m}

\software{{\tt astropy} \citep{astropy2013,astropy2018}, {\tt astrometry.net} \citep{lang2010}}



\begin{thebibliography}{} 

\bibitem[Astropy Collaboration et al.(2013)]{astropy2013} Astropy Collaboration, Robitaille, T.~P., Tollerud, E.~J., et al.\ 2013, \aap, 558, A33

\bibitem[Astropy Collaboration et al.(2018)]{astropy2018} Astropy Collaboration, Price-Whelan, A.~M., Sip{\H{o}}cz, B.~M., et al.\ 2018, \aj, 156, 123

\bibitem[Lang et al.(2010)]{lang2010} Lang, D., Hogg, D.~W., Mierle, K., et al.\ 2010, \aj, 139, 1782

\end{thebibliography}
\end{document}